\documentclass[twocolumn]{aastex631}

\usepackage{physics,amsmath,amssymb,bbold,amsthm,mathrsfs}
\usepackage{xcolor,url}

\usepackage{xspace}

\newcommand{\mdot}{{$\dot{M}\,\mathrm{yr}^{-1}$}\xspace}
\newcommand{\Change}[1]{{\color{red}\bf #1}}
\renewcommand{\Change}[1]{#1}

\newcommand{\mki}{
  Kavli Institute for Astrophysics and Space Research, 
  Massachusetts Institute of Technology , 77 Massachusetts  Ave., 
  Cambridge, MA 02139, USA
}

\begin{document}

\title{A Time-Dependent Spectral Analysis of $\gamma$ Cassiopeiae}

\author[0000-0003-2602-6703]{Sean J. Gunderson} \thanks{\url{seang97@mit.edu}}\affiliation{\mki}

\author[0000-0002-3860-6230]{David P.\ Huenemoerder}\affiliation{\mki}

\author[0000-0002-5967-5163]{Jos\'e M. Torrej\'{o}n}
\affiliation{Instituto Universitario de F\'{\i}sica Aplicada a las Ciencias y las Tecnolog\'{\i}as, Universidad de Alicante, E-
03690 Alicante, Spain}

\author[0000-0003-1898-4223]{Dustin K. Swarm}
\affiliation{Department of Physics and Astronomy, University of Iowa, Iowa City, IA, USA}

\author[0000-0003-3298-7455]{Joy~S.\ Nichols}
\affiliation{Harvard \& Smithsonian Center for Astrophysics, 60 Garden St., Cambridge, MA 02138, USA}

\author[0000-0002-1131-3059]{Pragati Pradhan}
 \affiliation{Embry Riddle Aeronautical University, Department of
 Physics \& Astronomy, 3700 Willow Creek Road Prescott, AZ 86301, USA}

\author[0000-0002-7204-5502]{Richard Ignace} \affiliation{Department
 of Physics \& Astronomy, East Tennessee State University, Johnson City, TN 37614 USA}

\author[0000-0003-4243-2840]{Hans Moritz Guenther}\affiliation{\mki}

\author[0000-0002-6737-538X]{A.~M.~T. Pollock}
\affiliation{Department of Physics and Astronomy, University of Sheffield, Hounsfield Road, Sheffield S3 7RH, UK}

\author{Norbert S.\ Schulz}\affiliation{\mki}

\begin{abstract}

We investigated the temporal and spectral features of $\gamma$ Cassiopeiae's X-ray emission within the context of the white dwarf accretion hypothesis. We find that the variabilities present in the X-ray data show two different signals, one primarily due to absorption and the other due to flickering like in non-magnetic cataclysmic variables. We then use this two-component insight to investigate previously un-reported simultaneous \textit{XMM} and \textit{NuSTAR} data. The model fitting results find white dwarf properties consistent with optical studies alongside a significant secondary, thermal source. We propose a secondary shock between the Be decretion disk and white dwarf accretion disk as the source. Finally, we analyzed a unique, low-count rate event of the \textit{XMM} light curve as potential evidence for the white dwarf encountering Be decretion disk structures.
\end{abstract}

\section{Introduction} \label{sec:intro}

\begin{figure*}
    \centering
    \includegraphics[width=\linewidth]{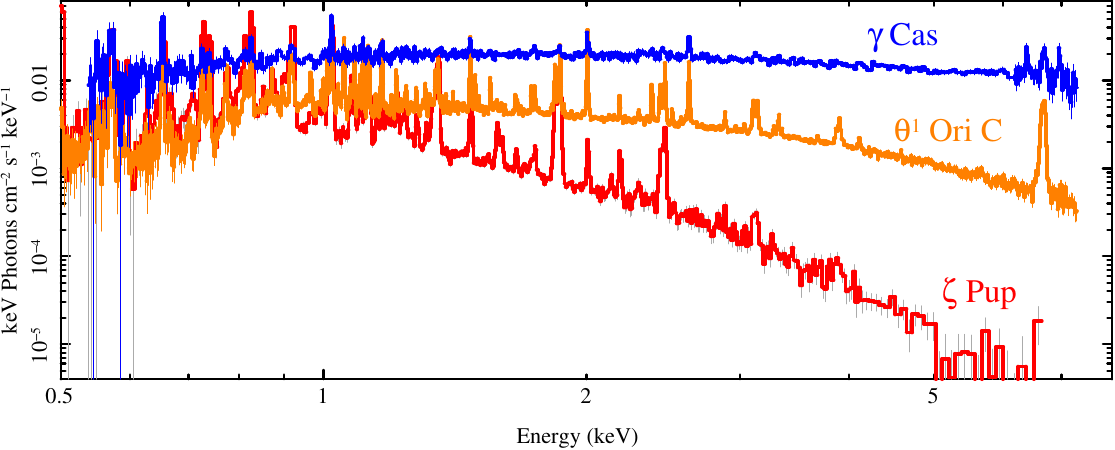}
    \caption{Comparison of the X-ray spectra of three different types of OB stars: $\zeta$ Puppis \citep[embedded wind shocks;][and citations therein]{Cohen2014,Huenemoerder2020}, $\theta^1$\,Ori\,C \citep[magnetically confined wind shocks,][]{Naze2014}, and $\gamma$\,Cas (unknown). All spectra are the \textit{Chandra} HEG+MEG plotted on the MEG grid. See Tables~\ref{tab:OtherObsIds} and \ref{tbl:oriobs} for information on  the $\zeta$\,Pup and $\theta^1$\,Ori\,C datasets used.}
    \label{fig:3StarCompare}
\end{figure*}

For the last 30 years, the X-ray emission of $\gamma$ Cassiopeia has been one the most difficult systems to understand. What makes the system so enigmatic is not its uniqueness, given that a number of analogues have been found (e.g., $\pi$\,Aqr \citep{Naze2017,Tsujimoto2023,Huenemoerder2024}, $\zeta$\,Tau \citep{Naze2024}, HD 110432 \citep{Tsujimoto2018}, HD 119682 \citep{Naze2020}, and HD 45314 \citep{Rauw2013}). The problem arises from the nature of its X-ray spectra compared to other OB stars. In Figure~\ref{fig:3StarCompare} we show $\gamma$\,Cas's \textit{Chandra} High Energy Transmission Grating (HETG) X-ray spectra compared to similar \textit{Chandra} observations of $\zeta$ Puppis \citep[a ``normal'' OB star;][]{Cohen2014,Huenemoerder2020} and $\theta^1$\,Ori\,C \citep[a strongly magnetic O star;][]{Naze2014}. What these spectra show is that $\gamma$\,Cas is not only exceptionally bright across all wavelengths but its X-ray source mechanism is much hotter than the others.

$\gamma$\,Cas is also known to be highly variable across all wavelengths \citep{Smith1995,Robinson2002,Smith2006,Lopes2010,Smith2019,Bore2020}, most notably in the optical from structure in the Be decretion disk and in X-rays from the unknown X-ray emission source. These variabilities occur at all timescales, from short time fluctuations within observations and longer period ``disk eruptions'' \citep{Rauw2022}. There are additional sub-variability structures that \citet{Smith2019} refer to as ``softness dips'' that are characterized as short periods of decreased soft counts and increased hardness ratio in the light curves. The combination of the hot X-ray emission and strong variability is why the X-ray emission mechanism has been difficult to decipher.

A number of models have been proposed, though, ranging from complex magnetic interactions to accretion onto a compact object; see \citet{Langer2020} for a detailed review of each of the proposed models. The magnetic interaction picture, more specifically the coupling between the Be star and its disk, has been the most ubiquitous model argued since its inception. However, \citet{Langer2020} argues that this magnetic picture is unlikely to produce the X-rays and is too difficult, if not impossible, to test.

On the accretion side, work has focused on the possibility of the compact object being a white dwarf \Change{\citep[WD;][]{Murakami1986,Tsujimoto2018,Gies2023}} based on the similarities in the spectra between $\gamma$\,Cas stars and cataclysmic variables. The accretion generated X-rays have a unique signature from their emission measure distribution \citep{Fabian1994} and fluorescent lines like Fe~K~$\alpha$ at 6.4\,keV \citep{Mukai2017}. Comparatively little work has been done to explain the temporal variations in the context of WD emission. Our goal is to do so to explain the variability in $\gamma$\,Cas. At the same time, we will support this temporal analysis by also looking at the spectral data in specific time slices and at different energy bands to form a complete temporal spectral picture of the system.

This paper is organized as follows. In \S~\ref{sec:DataReduction}, we give details to the X-ray observations we use and the data reduction methods. The Observation IDs of all data sets used in this paper are given in Appendix~\ref{sec:ObsIDs}. In \S~\ref{sec:Variability}, we analyze the types of variability that $\gamma$\,Cas could be undergoing. In \S~\ref{sec:SpectralFitting}, we fit simultaneous \textit{XMM} and \textit{NuSTAR} data and make time-sliced spectra to investigate light curve features. Finally, in \S~\ref{sec:Conclusions} we give our conclusions and proposed future work.

\section{Data Reduction}\label{sec:DataReduction}

The table of observations used in our analysis are given in Table~\ref{tab:ObsIds}. The \textit{Chandra} data was reprocessed with the standard pipeline in \textsc{ciao} version 4.15 \citep{Fruscione2006}. \textit{NuSTAR} data was processed with the standard pipeline in the \textsc{heasoft} version 6.33 tools provided by \textsc{sciserver} \citep{SciServer2020}. For \textit{XMM}, we used data products provided by the Pipeline Processing System and processed the available Observation Data Files into spectral data products using \textsc{sas} version 20.0.0 \citep{Gabriel2004}.

A primary interest of this work is the time-dependent spectral characteristics of $\gamma$\,Cas's X-ray data. Spectra were thus extracted with varying time lengths, ranging from 60\,s sub-intervals over an observation to the entire observation time. Each individual spectrum was given the same processing as appropriate for their respective observatory.

Light curves are presented throughout this work. In addition to the standard count rate light curves, we utilize fluxed light curves. To generate these, we used the unfolding principle implemented in the \textsc{isis} software to flux-correct the counts spectra. For details on the limits and background of this method, see \citet[in prep.]{Gunderson2024}. For a specified time binning, we extracted spectra in those intervals. These spectra were then unfolded and their flux was fit in the respective bands, generating a fluxed light curve with the same time binning as the extracted spectra.


\section{Variability Classification}\label{sec:Variability}

The first question we address is whether the short term variability in $\gamma$\,Cas is periodic or stochastic. In the context of the WD accretion hypothesis, either are possible based on whether the WD has a strong magnetic field or not. If the magnetic field is strong enough to control the accretion, then the accretion will be concentrated to spots on the poles of the surface that then rotate in-and-out of our view \citep{Mukai2015,Mukai2017,Lopes2019}. On the other hand, if the magnetic field is too weak to direct the flows, allowing material to accrete along the equator in a boundary layer, the variability will be stochastic. The stochasticity is better described by the observed phenomena called ``flickering'' \citep{Bruch1992,Balman2012}. In this situation, the change in the X-rays is caused by fluctuation in the inner edge of the accretion, though the exact mechanism is not known \citep{Bruch1992}.

\begin{figure*}
    \centering
    \includegraphics[width=\linewidth]{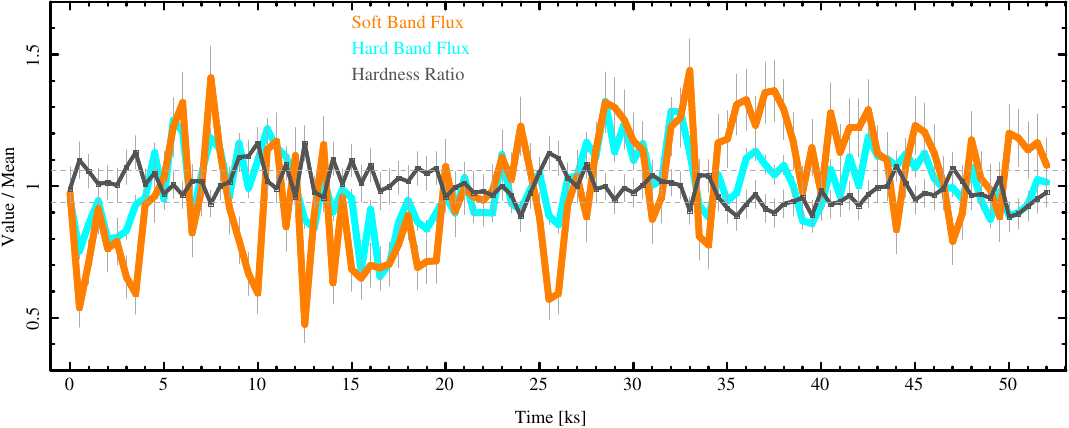} 
    \caption{Normalized light curves in the hard band ($1.7$--$7.0\,$\AA, cyan),
    soft band ($12$--$25\,$\AA, orange), and the hardness
    ratio defined as $(H-S)/(H+S)$.  The $1\sigma$ boundary of the hardness ratio is shown in black dashed lines. \Change{For reference, the mean, minimum, and maximum values before normalization were 
    $H: 110\,(72,\,146)$, 
    $S: 18\,(8.6,\, 26)$ 
    (both in units of 
    $\mathrm{10^{-12}\,erg\,cm^{-2}\,s^{-1}}$), 
    $\mathrm{HR}: 0.63\, (0.72, \,0.84).$}}
    \label{fig:BandHRLC}
\end{figure*}

An example of $\gamma$ Cas's X-ray mean-normalized light curve is given in Figure~\ref{fig:BandHRLC} using its \textit{Chandra} data. Specifically, we show the light curve for the hard band (1.7 -- 7.0\,\AA, cyan) and soft band (12 -- 25\,\AA, orange). We separate these two bands to highlight the complexity of $\gamma$\,Cas's variability and that there are two different signals within the light curve. The source of the two signals will be discussed later; here we want to classify the periodicity of the composite signal of these two bands.

\begin{figure*}
    \centering
    \includegraphics[width=\linewidth]{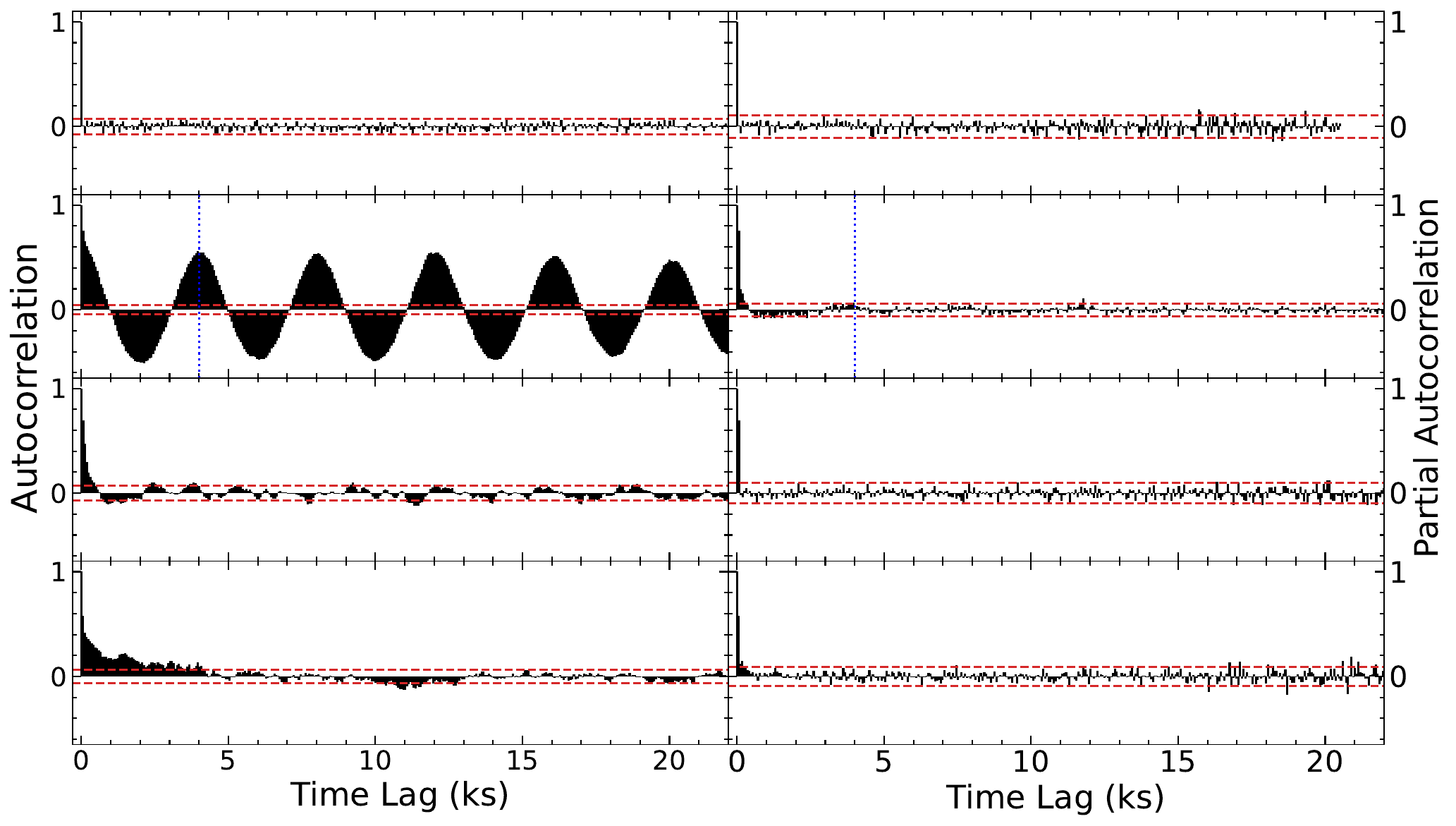}
    \caption{Autocorrelation (left column) and partial autocorrelation (right column) of the light curves of four systems of different variabilities. Capella: no intrinsic variability (top row). Ex Hya: periodic variability (second row; period marked with the dotted blue line). SS Cyg: stochastic variability (third row). Bottom row is $\gamma$ Cas, whose variability is not a priori classified. Red dashed lines in each plot mark the significance limit for the correlogram, defined as $\sigma_R = 1/\sqrt{N}$, where $N$ is the number of time bins.}
    \label{fig:AutoCorrPlots}
\end{figure*}

To \Change{broadly characterize the variability} of $\gamma$\,Cas, we will use the autocorrelation function
\begin{equation}
    R_{rr}(t_k) = \sum_{i=1}^n r(t_i)r(t_i-t_k),
\end{equation}
where $r(t_i)$ is the observed count rate in the time bin $t_i$. Autocorrelation is a versatile tool that can determine general correlations, periodicity, and the period of a time series data. In order to classify the periodicity of $\gamma$ Cas, we need examples of correlograms for different kinds of variability. Three stars are used: Capella, EX Hya, and SS Cyg. The \textit{Chandra} HETG ObsIDs of these objects are given in Table~\ref{tab:OtherObsIds}. These observations were processed in the same pipeline as described in \S~\ref{sec:DataReduction}. The stars' correlograms are given in the left column of Figure~\ref{fig:AutoCorrPlots} with 60\,s bins for all of them.

Three cases of variability are covered by our three control stars. Capella is our example of a non-variable system as it shows no variations on timescales shorter than an observation time \citep{Marshall2021}. The top row of Figure~\ref{fig:AutoCorrPlots} shows that there is no timescale at which the data shows significant variability. Here ``significant'' refers to a time-lag $t_k$ whose autocorrelation value $R_{rr}(t_k)$ is larger than the uncertainty (marked as red dashed lines in Figure~\ref{fig:AutoCorrPlots}) defined as $\sigma_R = 1/\sqrt{N}$, where $N$ is the number of time bins. Next, EX Hya is an intermediate polar \citep{Mukai2023} whose 67 minute period in the X-rays has been known for decades \citep{Heise1987}. As such, EX Hya is an example of a periodic signal in its correlogram in the second row of Figure~\ref{fig:AutoCorrPlots}. Note that periodic signals should decay toward the end of the observation, but we have restricted all the observations to the same 20\,ks window for readability. Finally, the CV SS Cyg covers the case of a stochastic signal through its flickering \citep{Albert2021}. The third row of Figure~\ref{fig:AutoCorrPlots} shows that SS Cyg has many significant timescales, but there is no distinct sinusoidal pattern.

The correlogram of the \textit{Chandra} observation of $\gamma$\,Cas is plotted in the last row of Figure~\ref{fig:AutoCorrPlots}. Of the three variabilities, $\gamma$\,Cas is most similar to that of SS Cyg's based on the similarity of their correlograms. However, $\gamma$\,Cas also shows more significant timescales that could be evidence of a weaker true periodicity. Regular autocorrelation makes this difficult to investigate because in a given period, $R_{rr}(t_k)$ includes the stochastic signal. \Change{To elucidate further the variability classification}, we can use the partial autocorrelation function
\begin{equation}
    \phi(t_k,t_k) = \frac{R_{rr}(t_k)-\sum_{i=1}^{k-1} \phi(t_{k-1},t_i)R_{rr}(t_{k-i})}{1-\sum_{i=1}^{k-1} \phi(t_{k-1},t_i)R_{rr}(t_{k-i})},
\end{equation}
\Change{where $\phi(t_k,t_i)=\phi(t_{k-1},t_i) - \phi(t_k,t_k)\phi(t_{k-1},t_{k-i})$ for $1\leq i \leq k-1$}. The advantage of the partial autocorrelation function is that a given time $t_k$ is only compared against itself. More explicitly, the partial autocorrelation of the light curve is the autocorrelation between $r(t_i)$ and $r(t_{i+k})$ with time bins $t_{i+1}$ to $t_{i+k-1}$ removed inclusively. As such, stochastic signals will be filtered out and reveal any periodic signals more clearly.

The correlograms of the partial autocorrelation are given in the right column of Figure~\ref{fig:AutoCorrPlots}. Going down the rows, we first see again that Capella has no significant signals reinforcing the lack of true variability. EX Hya shows only significant signals around its 67 minute period (marked with a dotted blue line). Moving to SS Cyg, we see that for a stochastic signal only the first immediate time bin is significant.\footnote{By ``first time bin'' we mean the first after $t=0$\,s, so $t=60$\,s in our case. By definition, the autocorrelation and partial autocorrelation functions are equal to 1 at $t=0$\,s.} This can be interpreted as only adjacent time bins, i.e. those only 60\,s apart based on our binning, are correlated with each other. Finally, comparing these to $\gamma$ Cas, we see that only the first non-zero time bin is significant, just as in SS\,Cyg. We can conclude that there is no periodic signal in $\gamma$\,Cas, i.e, both the soft and hard wavebands vary stochastically. 

Returning to the the light curves in Figure~\ref{fig:BandHRLC}, we can note that the variations in the two bands are sometimes correlated but at other times uncorrelated. These were noted by \citet{Smith2019} and further analyzed by \citet{Rauw2022}, who attributed it to changes in emission and absorption from ``canopies" of gas in their magnetic shearing model. However, we argue that the variability in both hard and soft bands is indicative of not just changes in absorption (which will primarily affect the softer wavelengths) but also two different emission sources.

In Figure~\ref{fig:BandHRLC}, we also include the mean normalized hardness ratio
\begin{equation}
    \mathrm{HR} = \frac{H-S}{H+S},
\end{equation}
where $H$ and $S$ are the hard ($1.7$--$7.0\,$\AA) and soft ($12$--$25\,$\AA) flux values respectively. The HR is driven by the changes in the soft flux, as was noted in prior works, but what we want to highlight is the quasi-independence of the hard flux to this absorption.

\begin{figure*}
    \centering
    \includegraphics[width=\linewidth]{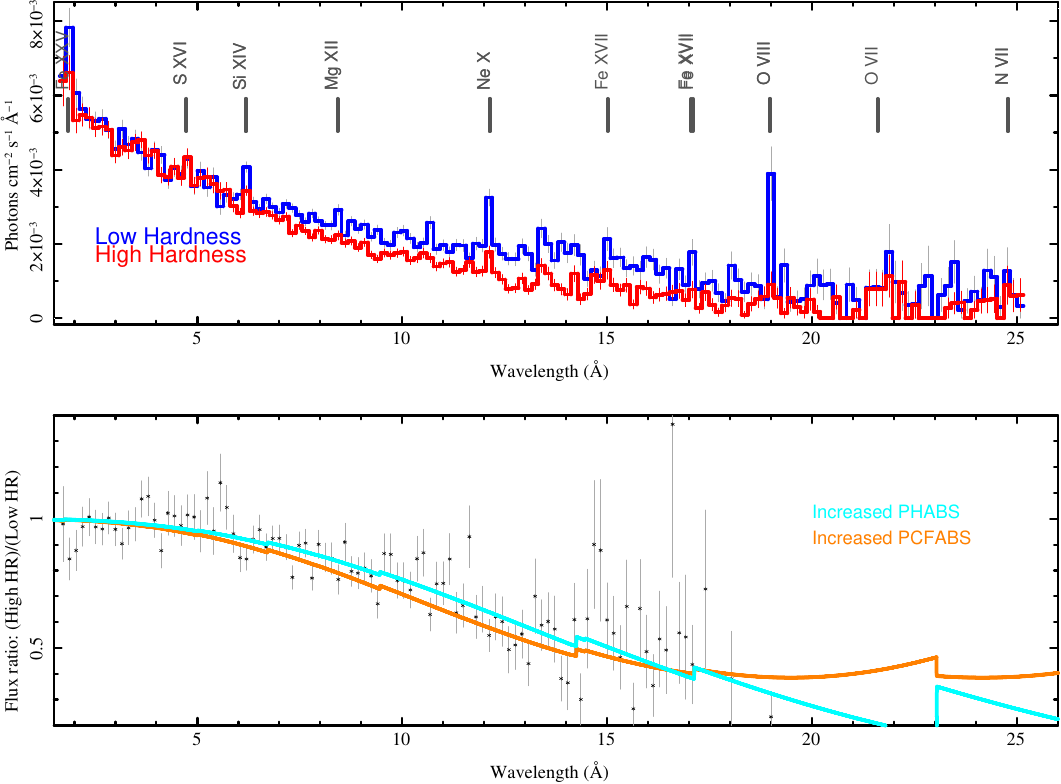} \caption{Top
    panel: spectra filtered by hardness ratio ranges.  The ``Low
    Hardness'' spectrum (blue) is from times when the HR was at least
    $1\sigma$ below the mean hardness ratio, and the ``High Hardness''
    from time when HR was at least $\sigma$ above the mean.  Bottom:
    the ratio of the high to low HR spectra (black points with error
    bars, for points with $>2\sigma$ significance).  The
    colored curves show absorption functions for slab absorption
    (``PHABS,'' cyan) and partial covering (``PCFABS,'' orange) for an
    increase in the column by $0.2\times10^{22}\,\mathrm{cm^{-2}}$
    from the nominal models, showing that the higher hardness state is consistent with increased absorption.}
    \label{fig:SpecRatioHR}
\end{figure*}

In the top panel of Figure~\ref{fig:SpecRatioHR} we show the HEG+MEG HETG spectra (on the MEG grid) of $\gamma$\,Cas in high (red) and low (blue) hardness states. We define these hardness states as the times when $\mathrm{HR}\gtrless\overline{\mathrm{HR}}\pm\sigma_\mathrm{HR}$, where $\overline{\mathrm{HR}}$ is the mean HR and $\sigma_\mathrm{HR}$ is the standard deviation. These two plots show the primary effect is absorption as the high HR state has less flux in the longer wavelengths. This is shown more explicitly in the bottom plot where the ratio of the high HR to low HR in every wavelength bin is plotted. Two absorption models plotted: simple slab absorption (``PHABS,'' cyan) and partial covering (``PCFABS,'' orange). Both these models use an increase in column density of $N_\mathrm{H}=0.2\times10^{22}\,\mathrm{cm^{-2}}$, assuming solar abundances \citep{Anders1989}.

\begin{figure}
    \centering
    \includegraphics[width=\linewidth]{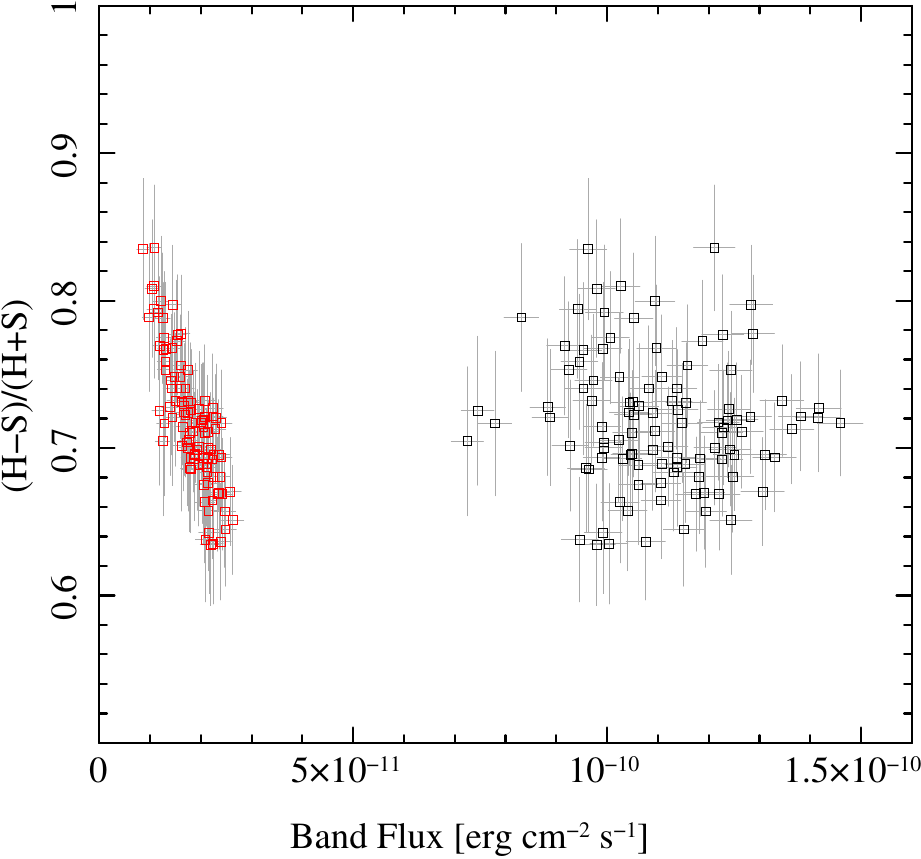} 
    \caption{Hardness ratios against band flux.  Points are shown twice, once against
    the soft flux ($12$--$25\,$\AA, left side, red), and
    again vs hard flux ($1.7$--$7.0\,$\AA, right side,
    black).  Note the strong correlation of the hardness ratio with
    soft flux, and lack of correlation with hard flux.}
    \label{fig:HRvsFlux}
\end{figure}

Despite the increase in column density, the ($\lambda<7$\,\AA) are unaffected by the absorption. Yet in Figure~\ref{fig:BandHRLC} the hard band is varying just as much as the soft band. Hence we argue that there is a second, true \textit{source} variability in $\gamma$\,Cas associated with its X-ray generation mechanism. This is more easily seen in Figure~\ref{fig:HRvsFlux} where we plot the HR against the soft and hard flux values in every bin. The correlation between HR and soft flux is plainly visible in the linear trend with the soft flux (red points). The hard flux (black points) on the other hand show no correlation. Instead, there is a random distribution of the hard flux over all HR. We argue that we are seeing true source variability in this case.

Given we are assuming the X-ray emission source is an accreting WD, this source variability is likely the flickering like in SS Cyg. More importantly, since there is no periodicity in the signals, we can argue against the \Change{intermediate polar} accretion mechanism. Thus we can conclude that an accretion disk is present and extends down to the surface, creating a boundary layer where the accretion X-rays are generated.

\section{Simultaneous Spectra: \textit{XMM} and \textit{NuSTAR}}\label{sec:SpectralFitting}

This work is not the first time that the individual exposures of the simultaneous \textit{XMM} and \textit{NuSTAR} observation (Obs ID's given in Table~\ref{tab:ObsIds}) has been used. The \textit{XMM} portion was used by \citet{Smith2019} for detailed investigation of the timing data. The \textit{NuSTAR} half was used by \citet{Tsujimoto2018} alongside a \textit{Suzaku} observation taken on 2011-07-13 to investigate the WD accretion hypothesis. The inclusion of the \textit{Suzaku} data was to constrain the low energy portion of $\gamma$\,Cas's emission, but we argue that observations taken at different times cannot be fit simultaneously because of $\gamma$\,Cas's variability between observation epochs \citep{Rauw2022}. By using the simultaneous \textit{XMM} and \textit{NuSTAR} data together, our model fitting is not subject to the epochal variations and will better describe the snapshot of the system.

\subsection{Average Spectra}

\begin{figure*}
    \centering
    \includegraphics[width=\linewidth, trim=40 0 0 0]{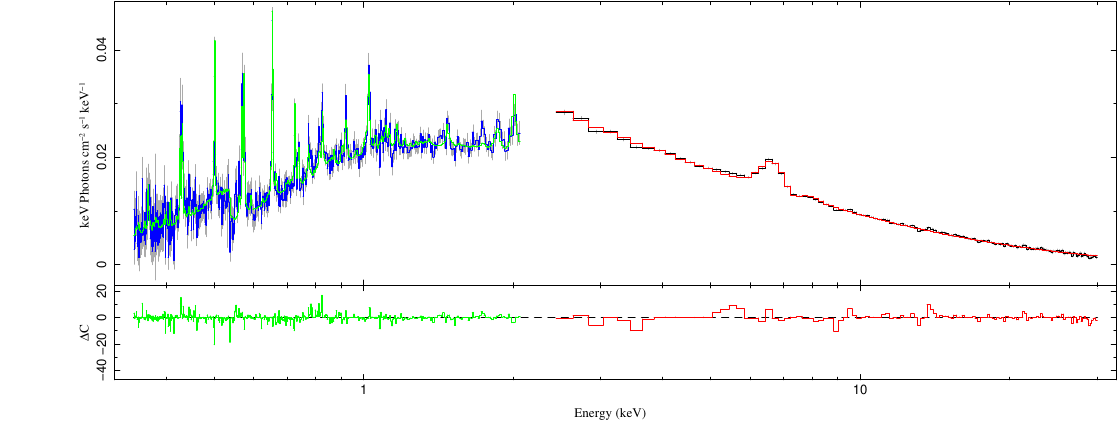}
    \caption{Simultaneous RGS (blue histogram) and NuStar (black histogram) $\gamma$ Cas model fit. The red histogram is the cooling flow and Gaussian line model applied to only the NuSTAR data. The green histogram is the cooling flow and power law DEM applied to only the RGS data.}
    \label{fig:RGSNuStargCas}
\end{figure*}

We will analyze the temporal variations in the spectra later in this section, but we first need to know the baseline, average properties of the spectra. For this, we will fit \textit{XMM}'s Reflection Grating Spectrometer (RGS) and \textit{NuSTAR's} Focal Plane Modules A and B (FPMA/B). While the \textit{XMM}'s EPIC pn (hereafter PN) or Metal Oxide Semi-conductor (MOS) cameras would provide additional counts in the high energy portion where \textit{NuSTAR} is sensitive, the high energy portion of the spectra is primarily continuum emission (see Figures~\ref{fig:RGSNuStargCas} or \ref{fig:DynamicSpectra} for example) beyond the Fe~K\,$\alpha$ complex. Additional photons would not significantly improve the signal-to-noise given the added computational overhead. The emission lines in the RGS spectra, on the other hand, provide better constraints on the differential emission measure (DEM) distribution while also adding the low energy spectrum we need.

When fitting the spectra of $\gamma$\,Cas and its analogues, it is common to use multi-temperature plasma models so as to fit both the H-like and He-like ions. H-like ion emissivity curves have long, high temperature tails that allow them to emit at much higher temperatures that are too hot for He-like emission (see figure 3 of \citealp{Gunderson2023} for an example of emissivity curves). \citet[][and citations therein]{Rauw2022} have applied such methods with multi-temperature plasma models that include $T < 2$\,keV components and a significant $T \geq 10$\,keV component. However, their interpretation of the different temperature components was not of two different sources.

In comparison, \citet{Hamaguchi2016} made the first direct interpretation of two emission sources based on \textit{Suzaku} data, specifically inferring polar accretion to be the source of the hot component and wind interactions between the Be star and WD as the soft source(s).

Our goal is to build off these multi-temperature, multi-source models to create a complete picture of the X-ray generation mechanism in $\gamma$\,Cas. Specifically, one model will be used to describe the WD accretion whereas the second source will be an unknown component that we aim to elucidate possible information on. As will be shown below, this second source is possibly the interaction between the WD's accretion disk and the surrounding Be decretion disk.

The first component is a spherically symmetric cooling flow to describe the accretion of material onto the WD, as was used by \citet{Tsujimoto2018}. The luminosity of the cooling flow is
\begin{equation}
    L(\lambda)d\lambda = \frac{5}{2}\frac{k_B\dot{M}}{\mu m_p}\int_{T_\mathrm{min}}^{T_\mathrm{max,1}}\frac{\Lambda_\lambda(T)}{\Lambda_\mathrm{Bolo}(T)}\left(\frac{T}{T_\mathrm{max,1}}\right)^\beta dTd\lambda,
\end{equation}
where $k_B$ is Boltzmann's constant, $\mu=0.67$ is the mean molecular weight, $m_p$ is the proton mass, $\dot{M}$ is the WD accretion rate, $T_\mathrm{min}$ and $T_\mathrm{max,1}$ are the lower and upper limits of the cooling flow's temperature distribution, $\Lambda_\lambda(T)$ is the emissivity of the plasma at the wavelength $\lambda$, and $\Lambda_\mathrm{bolo}(T)$ is the bolometric emissivity \citep{Fabian1994}. \citet{Dirk2005} found better fits to the spectra of CVs when a power law weighting was added to cooling flow models, so we have included a similar weighting with a power $\beta$. This component is accompanied by a partial covering fraction absorption model
\begin{equation}
    M_\mathrm{pcfabs}(\lambda) = f \mathrm{e}^{-n_{\mathrm{H},1} \sigma(\lambda)}+(1-f),
\end{equation}
where $f$ is the fraction of the boundary layer that is covered.

The free parameters of the cooling flow are the maximum temperature $T_\mathrm{max,1}$, mass-accretion rate $\dot{M}$, and the power law slope $\beta$. The minimum temperature $T_\mathrm{min}$ was frozen at $T_\mathrm{min}=1$\,MK. When both the minimum and maximum temperatures were allowed to be free, we found that the model fitting became highly degenerate. When one parameter would be constrained, the other would not be. Freezing the minimum temperature to its lower bound thus served to constrain the model parameter fitting.


We also included a Gaussian line model centered on the Fe~K\,$\alpha$ line (6.4\,keV) with an assumed, frozen width of $\sigma_{\mathrm{K}\,\alpha} = 0.05$\,keV. There are physically important processes that the Fe~K\,$\alpha$'s presence signifies \citep{Mukai2015,Mukai2017,Hayashi2018}, but these processes are not within the scope of this work. As such, the flux in the Fe~K\,$\alpha$ is a free parameter for improving the fit statistic and better constraining the other parameters in our model.

The second model component we include is for the unknown second source in the system. While multi-temperature plasma models, i.e., multi-temperature \textsc{apec} models, can adequately fit the data, they introduce many more parameters. To prevent our parameter space from growing too large, we used a power law DEM
\begin{equation}
    \frac{d\mathrm{EM}}{dT} = D_0(d)\left(\frac{T}{T_\mathrm{max,2}}\right)^\gamma,
\end{equation}
where $T_\mathrm{max,2}$ is the maximum temperature of the second source, $\gamma$ is the slope of the distribution, and $D_0(d)$ is a distance $d$ dependent normalization, to describe a multi-temperature source. This power law DEM is constructed through a weighted sum of isothermal plasma models; for more details on the model, see \citet{Huenemoerder2020}. This component will have its own absorption model applied to it, specifically a simple slab absorption
\begin{equation}
    M_\mathrm{phabs}(\lambda) = \mathrm{e}^{-n_\mathrm{H,2}\sigma(\lambda)}.
\end{equation}
In our modelling, we allowed the two column densities $n_\mathrm{H,1}$ and $n_\mathrm{H,2}$ to be independent. This is again a choice based on our lack of knowledge of the secondary source. It may be the case that the two sources are geometrically related such that they share the same absorption, but those are details we aim to find.

\citet{Pradhan2023} found that the emission lines in $\gamma$\,Cas had widths of order 1000\,km\,s$^{-1}$, so we used thermally broadened profiles. The first step in our model fitting was a simple model fit to find the appropriate line broadening for the two models to improve the fit statistic. These came out to $v_1 = 697$\,km\,s$^{-1}$ for the cooling flow model and $v_2=661$\,km\,s$^{-1}$ for the power law DEM. These velocities were frozen in our full Markov Chain Monte Carlo model fits.

The final parameters of our model are the abundances of different elements. Elements whose emission lines are resolved in the RGS spectra had freely fit abundances. Those that were not detected in the RGS had their abundances frozen at unity, relative to \citet{Anders1989}. Both of our model components are plasma models with their own abundance parameters. We set these parameters equal to each other so that we have only one set of abundance parameters describing the X-ray emitting plasma. This is not necessarily a requirement and could be relaxed to have two independent plasma models, like the absorption components. However, whatever the second source is, if it is within the $\gamma$\,Cas system, it should have a similar chemical composition.

The final modelling choice of note is how we applied the model to the data. Our assumption has been that the second source is only seen in the softer portion of the spectra, so it should not provide any flux to the harder portion. To account for this, we applied the power law DEM to only the RGS. The RGS model was then the sum of the power law DEM and the cooling flow with their respective absorption components.

The FPMA/B spectra only had the cooling flow applied to them along with an overall cross-calibration constants. \citet{Madsen2017} calculated cross-calibration constants for the ensemble of operating X-ray observatories at the time of their writing, but while there is a value given for comparing \textit{XMM} to \textit{NuSTAR}, the RGS was not included. The RGS is well-calibrated against the PN camera\footnote{\url{https://xmmweb.esac.esa.int/docs/documents/CAL-TN-0052.pdf}} to agree up to $\mathrm{RGS}/\mathrm{PN} = 0.95$ whereas \citet{Madsen2017} gives the PN camera to be in agreement with the FPMA up to $\mathrm{PN}/\mathrm{FPMA}=0.88-0.90$. We propagated these cross-calibration constants to determine that the RGS-to-NuSTAR constant should be $\mathrm{RGS}/\mathrm{FPMA} = 0.84$. However, we found that this value did not match the simultaneous data. As such, we allowed a cross-calibration constant to be free during initial fits, finding $\mathrm{RGS}/\mathrm{FPMA} = 0.70$.

We used the S-Lang \textsc{emcee} package in \textsc{isis} to fit our model using Markov Chain Monte Carlo methods. The \Change{Markov chain} was run long enough for the chain to stabilize, after which parameter contours were drawn. We assumed Poisson statistics and used a Cash fit statistic \citep{Cash1979}. The results of the \Change{Markov chains} are given in Table~\ref{tab:AvgFitResults} while the best fit is given in Figure~\ref{fig:RGSNuStargCas}.

\begin{deluxetable}{llc}
    \tablecaption{\textit{XMM}+\textit{NuSTAR} average spectra best fit parameters with $1\sigma$ uncertainties. \label{tab:AvgFitResults}}
    \tablehead{
        \colhead{Model Component} & \colhead{Parameter} & \colhead{Value}
    }
    \startdata
    Cooling flow & $\dot{M}$ ($10^{-10}$\,\mdot) & $3.38_{-0.77}^{+0.29}$ \\
     & $\beta$ & $0.30_{-0.23}^{+0.08}$ \\
     & $T_\mathrm{max,1}$ (MK) & $288.4_{-6.6}^{+27.8}$ \\
     & $v_\mathrm{turb}$ (km s$^{-1}$) & 697 \\
     & $A$(C) & $0.17_{-0.05}^{+0.06}$ \\
     & $A$(N) & $1.54_{-0.18}^{+0.61}$ \\
     & $A$(O) & $0.37_{-0.03}^{+0.38}$ \\
     & $A$(Ne) & $0.92_{-0.13}^{+0.13}$ \\
     & $A$(Mg) & $0.64_{-0.25}^{+0.24}$ \\
     & $A$(Fe) & $0.33_{-0.01}^{+0.01}$ \\
     & $n_\mathrm{H}$ $10^{22}\,\mathrm{cm}^{-2}$ & $0.22_{-0.01}^{+0.01}$ \\
     & $f$ & $0.92_{-0.03}^{+0.02}$ \\
     \hline
     DEM & Norm ($10^{-17}$ VEM / $4\pi d^2$) & $8.794_{-2.076}^{+0.890}$ \\
     & $\gamma$ & $-1.91_{-0.27}^{+0.27}$ \\
     & $T_\mathrm{max,2}$ (MK) & $13.5_{-12.0}^{+40.2}$\\
     & $n_\mathrm{H}$ $10^{22}\,\mathrm{cm}^{-2}$ & $0.06_{-0.02}^{+0.04}$ \\
    \enddata
\end{deluxetable}

The first result of note is the mass-accretion rate. Our best fit value is twice as large as \citet{Tsujimoto2018} found and is not consistent at the 68 percent level. This can be attributed to our use of the RGS instead of \textit{Suzaku}. One may question if the addition of the power law DEM skewed the parameter space of the cooling flow, in particular the accretion rate which acts as a normalization, yet if this was the case one would expect $\dot{M}$ to go down from the added flux of the DEM model. As such, this higher $\dot{M}$ is more accurately describing the average rate of the observation. It also highlights the need to use simultaneous data for such highly variable systems.

A common argument against the accreting WD hypothesis is that there is not enough mass within the Be decretion disk to support the calculated accretion rates. As a test, we can estimate the amount of accretion that the disk can support from a massive body moving through it using Bondi-Hoyle accretion
\begin{equation}
    \dot{M}_\mathrm{BH}\approx \pi\frac{\rho G^2 M^2}{c_s^3},
\end{equation}
where $G$ is the gravitional constant, $\rho$ is the density of the accreting material, $M$ is the WD mass, and $c_s$ is the sound speed of the gas around the accreting body.

First, we need a mass of the WD. We can get this from the max temperature of the cooling flow. In a non-magnetic CV, only half the gravitational potential energy is converted to kinetic energy \citep{Ezuka1999,Yu2018}, so the maximum temperature is
\begin{equation}
    k_B T_\mathrm{max,1} = 8\left(\frac{M}{0.5 M_\odot}\right)\left(\frac{R_\mathrm{WD}}{10^9 \mathrm{cm}}\right)^{-1}\,\mathrm{keV} \label{eq:kTtoM}.
\end{equation} 
Combining this with the mass-radius relation in for WDs
\begin{equation}
    R_\mathrm{WD} = 0.78\times10^9\left(\left(\frac{1.44 M_\odot}{M_\mathrm{WD}}\right)^{2/3} - \left(\frac{M_\mathrm{WD}}{1.44 M_\odot}\right)^{2/3}\right)^{1/2}\,\mathrm{cm},\label{eq:MassRadiusWD}
\end{equation}
\citep{Ezuka1999}, we can estimate the WD mass to be $M_\mathrm{WD}=0.93\,M_\odot$. Other relevant parameters of the $\gamma$\,Cas system are given in Table~\ref{tab:WDProps}, including quantities found in the literature and calculated here.

\begin{deluxetable}{lc}
    \tablecaption{Properties of stars in $\gamma$\,Cas. \label{tab:WDProps}}
    \tablehead{
        \colhead{Parameter} & \colhead{Value}
    }
    \startdata
    $M_*$ ($M_\odot$)\;\tablenotemark{a} & 13\\
    $R_*$ ($R_\odot$)\;\tablenotemark{b} & 10\\
    $T_\mathrm{eff}$ (kK)\;\tablenotemark{b} & 25\\
    $\log \rho_0$ (dex\,g\,cm$^{-3}$)\;\tablenotemark{c} & $-10.48_{-0.12}^{+0.10}$\\
    $M_\mathrm{WD}$ ($M_\odot$)\;\tablenotemark{d} & $0.93_{-0.01}^{+0.04}$\\
    $R_\mathrm{WD}$ ($R_\oplus$)\;\tablenotemark{d} & $0.94_{-0.05}^{+0.01}$\\
    $P$ (days)\;\tablenotemark{a} & $203.523\pm0.076$\\
    $a_*$ ($R_*$)\;\tablenotemark{d} & $1.93_{-0.02}^{+0.08}$\\
    $a_\mathrm{WD}$ (AU)\;\tablenotemark{d} & $1.632_{-0.001}^{+0.002}$
    \enddata
    \tablenotetext{a}{\citet{Nemravova2012}}
    \tablenotetext{b}{\citet{Sigut2007}}
    \tablenotetext{c}{\citet{Klement2017}}
    \tablenotetext{d}{Derived in this work.}
\end{deluxetable}

Our calculated mass value is similar to the $M=0.98\,M_\odot$ value \citet{Nemravova2012} found using radial velocity measurements, adding further validation to our model fits, but note that there is a systematic, unquantified uncertainty in our mass estimation due to the flickering. During subsequent observations, the overall flickering amplitude of the system may be higher or lower, changing the average spectra and subsequently the model parameters, such as those found by \citet{Tsujimoto2018} when using non-simultaneous data.

Next we need the density of the decretion disk material at the radius at which the WD orbits. The density of Be decretion disks in the viscous limit \citep{Klement2017} can be modeled by
\begin{equation}
    \rho_\mathrm{DD}(r,z) = \rho_0 \left(\frac{r}{R_*}\right)^{-\alpha} \mathrm{exp}\left(-\frac{z^2}{2H^2}\right),\label{eq:BeDiskDensity}
\end{equation}
where $\rho_0$ is the central density of the disk near the surface of the Be star, $r$ is the disk radius, $R_*$ is the radius of the Be star, $z$ is the height above or below the disk. The scale height of the disk $H$ is also dependent on the radius from the star:
\begin{equation}
    H(r)=H_0 \left(\frac{r}{R_*}\right)^{3/2},\label{eq:ScaleHeight}
\end{equation}
where $H_0 = c_s R_*/v_\mathrm{orb}$ is the scale height at the base of the disk and $v_\mathrm{orb}$ is the orbital velocity of the disk material. The parameter $\alpha$ describes how fast the density drops in the disk, which varies from star to star. In steady-state, the exponent is $\alpha=3.5$ \citep{Klement2017}, which we adopt here.

The radius of interest corresponds to the WD's semi-major axis $a_\mathrm{WD}$, which we can get from the ephemeris from \citet{Nemravova2012} and our calculated WD mass, all of which are given in Table~\ref{tab:WDProps}. Note though that this will not be the radius within the decretion disk, which is measured from the Be star. Since the WD's mass is not neglible, the system's center mass is $a_* = (M_\mathrm{WD}/M_*)a_\mathrm{WD} = 1.93_{-0.02}^{+0.08}\,R_*$ from the Be star, so the radius within the Be decretion disk is $r=a_\mathrm{WD}+a_*$. \citet{Klement2017} estimated that the central density of the disk in $\gamma$\,Cas to be $\log \rho_0 = -10.48_{-0.12}^{+0.10}$\,dex\,g\,cm$^{-3}$. Assuming that the WD orbits within the $z=0$ central portion of the decretion disk, the density at $r=a_\mathrm{WD}+a_*$ is $\log\rho\approx-15.99_{-0.05}^{+0.08}$\,dex\,g\,cm$^{-3}$.

Finally, the sound speed of the gas $c_s = \sqrt{5 k_BT/3 \mu m_p}$ around this point is dependent on the temperature of the disk. Be disks can be well described as isothermal and are dependent on the temperature of the radiation from the Be star. \citet{Sigut2009} found that for most Be stars, the decretion disk temperature is approximately $T=0.6 T_\mathrm{eff}$, which gives sound speeds of $c_s=17.57$\,km\,s$^{-1}$. Putting these all together, we find that the Bondi-Hoyle accretion rate is then $\dot{M}_\mathrm{BH}=1.43_{-0.18}^{+0.28}\times10^{-8}\,M_\odot\,\text{yr}^{-1}$.

Such a high value for the Bondi-Hoyle accretion rate shows two things. First and foremost, that there is more than enough mass within the Be decretion disk at any time to fuel the WD accretion. This is especially true given that we have not accounted for the WD's gravitational field extending the region it can accrete from. From Equation~\eqref{eq:BeDiskDensity}, the density is strongly sloped in radius, so with almost a solar mass in gravitational potential, the WD is able to pull material from inner regions with higher density that would increase $\dot{M}_\mathrm{BH}$. Secondly, it affirms our interpretation of the timing data as evidence of an accretion disk. Disk accretion is significantly less efficient than spherical accretion due to the need of the material to shed angular momentum.

While this Bondi-Hoyle accretion is not what is fueling the X-rays we see, it is still a significant process in the system. The WD will be collecting gas at this rate, which may explain our absorption model results. Both the cooling flow and power law DEM model components converged to relatively little absorption, especially the power law DEM. At the same time, though, the cooling flow has a much larger covering fraction at $f=0.92$ than \citet{Tsujimoto2018} found. Such a large covering fraction can be difficult to explain for a non-magnetic CV since we need to cover the entire equatorial region of the WD, but the collecting gas predicted by the Bondi-Hoyle accretion rate provides an explanation.

The WD will attempt to collect $10^{-8}\,M_\odot\,\text{yr}^{-1}$ worth of material, but this gas will need to join up with the accretion disk, which only loses $10^{-10}\,M_\odot\,\text{yr}^{-1}$. Assuming the accretion disk is steady state, it will only replenish this latter amount, causing an accumulation of mass in a halo around the WD. We will refer to this spherical distribution as an accretion halo to distinguish it from the accretion disk. This \Change{accretion halo}, based on its density, would provide the covering fraction we found in our X-ray model fits since it would cover the entire WD but not be entirely opaque, giving us the $f=0.92$ covering fraction.

The process of the \Change{accretion halo} material joining the \Change{accretion disk} will not be done smoothly due to the difference in angular momentum. The merging of the different materials will form a shock that we propose is the second source within the $\gamma$\,Cas system. From the power law DEM model, the maximum temperature $T_\mathrm{max,2} = 13.5$\,MK is the highest temperature of the circular distribution of shocks around the accretion disk. $T_\mathrm{max,2}$ is significantly lower than the accretion shock's temperature of $T_\mathrm{max,1}=288.4$\,MK, so one may question whether the power law DEM model is providing any significant flux, or in other words, that this \Change{accretion-halo-accretion-disk} shock is not bright enough to see above the accretion emission.

\begin{figure*}
    \centering
    \includegraphics[width=\linewidth, trim=40 0 0 0]{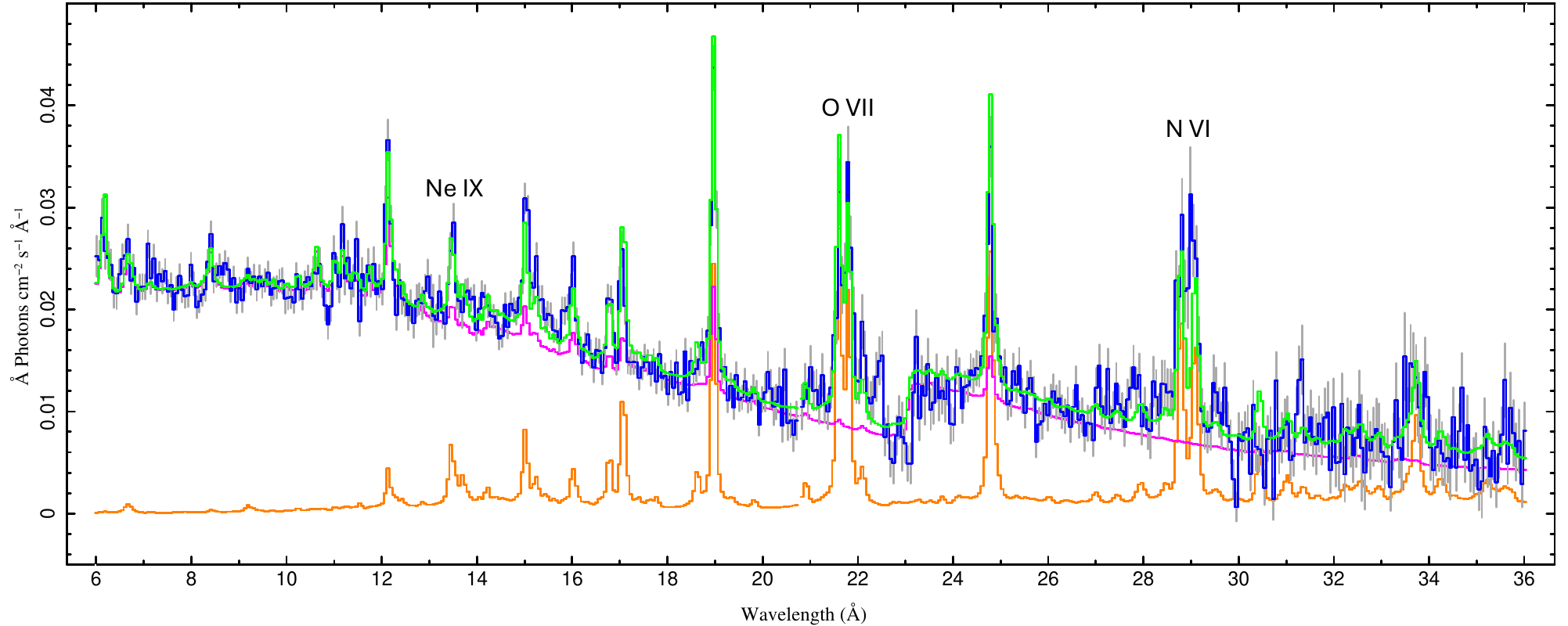}
    \caption{RGS $\gamma$ Cas model fit (total; green) with the model components: cooling flow (purple) and power law DEM (orange). Data (blue) is binned by a constant factor of 6. See Figure~\ref{fig:RGSNuStargCas} for residuals of the total model fit.}
    \label{fig:RGSgCas}
\end{figure*}

In Figure~\ref{fig:RGSgCas}, we plot just the RGS spectra and its model fit to show that there is significant He-like emission line flux, which we would not expect from the extreme high thermal temperatures of the cooling flow and its slight positive emission measure slope. The power law DEM, however, is at a temperature we would expect significant He-like emission, which we see in the labeled lines in Figure~\ref{fig:RGSgCas}. The power law DEM is not providing any significant continuum flux but is a line \Change{source}.

Before we move on, it worth considering alternative interpretations of the secondary source. The best-fit parameters of the power law DEM are similar to those found from fitting embedded wind shock spectra \citep{Huenemoerder2020}. As an early B star, $\gamma$\,Cas would be X-ray bright \citep{Pradhan2023}, so the He-like emission could be from embedded wind shocks, like \citet{Hamaguchi2016} argued. However, as will be discussed in the following section, there is a long-lived absorption event that challenges this interpretation.

\subsection{Time-dependent Spectra: Deep Absorption Events}

Since we are assuming all of the X-rays are coming from a single point source in the disk\footnote{Compared to the primary Be star and the decretion disk, the WD and its \Change{accretion disk} would be significantly smaller and effectively still a single point.}, there are broader implications for the features previously noted in the data. First and foremost are the softness dips in the light curves. Prior work by both \citet[][and citations therein]{Smith2019} and \citet{Hamaguchi2016} interpreted these as being due to absorption. The latter work by \citet{Hamaguchi2016}, in the context of the WD accretion model, provided a simple geometric argument that the wind clumps that make up the Be star's wind could occult the WD and be the source of the absorption.

\begin{figure*}
    \centering
    \includegraphics[width=\linewidth]{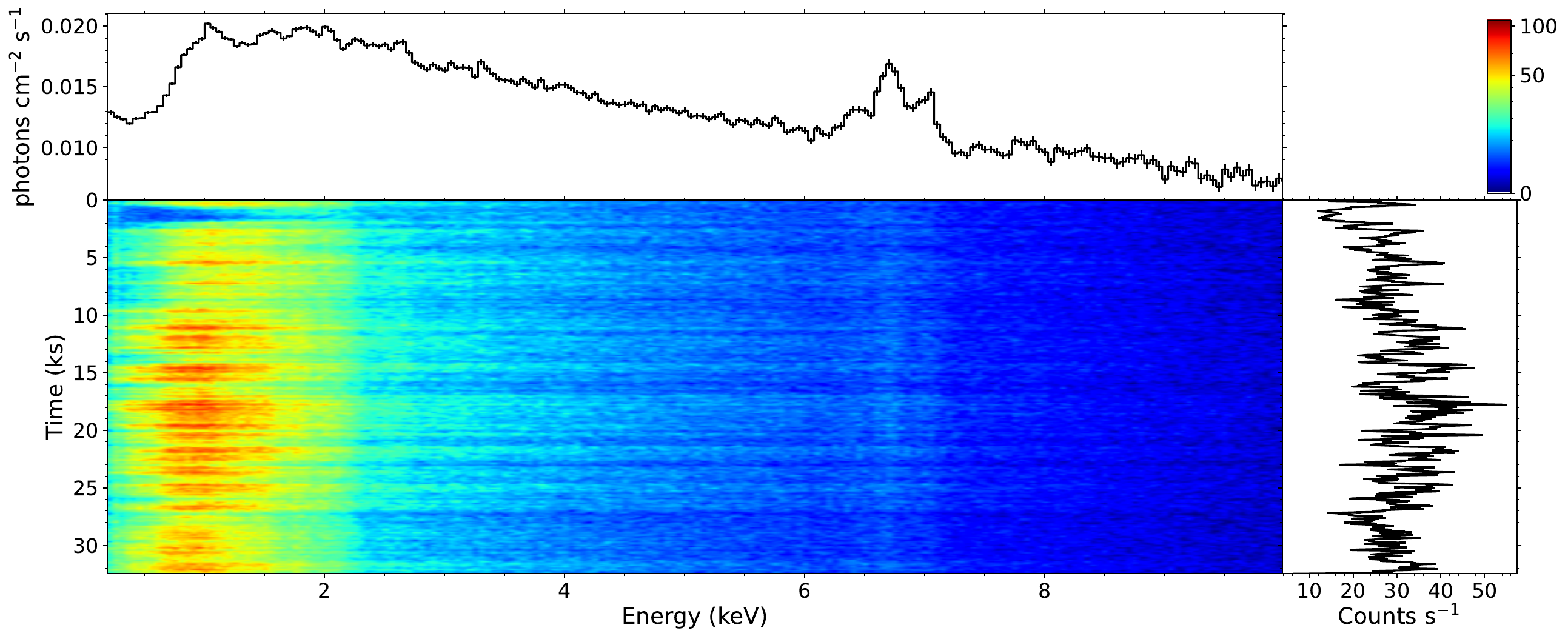}
    \caption{$\gamma$ Cas \textit{XMM} dynamic spectrum. Top: \textit{XMM} PN spectra binned by a constant factor of 10. Bottom left: Count map of the \textit{XMM} PN camera observation binned by a constant factor of 10 in energy and 60\,s in time. Colors correspond to the amount of counts in the bin with the legend in the top right corner. Bottom right: Corresponding count rate light curve of the observation binned at 60\,s.}
    \label{fig:DynamicSpectra}
\end{figure*}

The \textit{XMM} data that we are using has a known event in its light curve where the count rates drop significantly, more so than the quiescent variability and for much longer. This event is from about 1-3~ks in the light curve in the bottom right panel of Figure~\ref{fig:DynamicSpectra}. \citet{Smith2019} classified the feature as a \Change{softness dip} as well, though this is perhaps too general of a name. The word ``dip'' evokes more of a short term signature, which the quiescent variability already shows. The event in question also appears to be unique compared to the overall variability. As such, we will refer to this feature as simply the ``Trough'' and defer giving it more appropriate names to after our analysis. The authors of \citet{Smith2019} analyzed the Trough in the context of the magnetic model, but this same event has not been explained in the context of the WD hypothesis.

We can note a few characteristics of the Trough. First, the Trough has a similar shape to an exoplanet transit, though we caution against calling it a transit directly. This specific shape can be more generally interpreted as a smooth, gradual process, unlike the usual low count absorption features seen in the light curve of Figures~\ref{fig:BandHRLC} and \ref{fig:DynamicSpectra} that are sharper in transition. Using similar terminology to transits, the Trough appears to have a U-shape while the usual absorption features have a V-shape. Secondly, the Trough is more long lived, lasting for between 1.5 -- 2\,ks. One may argue that the identified \Change{softness dips} in the light curves analyzed by \citet{Smith2019} also show a U-shaped signatures, but only this one \textit{XMM} observations shows the most distinctive U-shape.

The flickering is still present throughout the Trough, making it difficult to note exact timings. There is also a drop during the egress that appears to prolong it. If we assume that the ingress and egress are actually symmetric, then we can estimate the ingress/egress times to be $t_\mathrm{ingress} = 0.42$\,ks while the bottom of the Trough lasts for $t_\mathrm{bot}=0.84$\,ks, so the best estimate for the total event time is 1.68\,ks.

This estimate also assumes that the loss in the counts happens evenly across all wavelengths. To examine this, we need a dynamic spectrum, which the total of Figure~\ref{fig:DynamicSpectra} is for the PN camera spectra of the \textit{XMM} observation we are analyzing. The top panel is the unfolded flux spectra of the PN binned by a constant factor of 10. The bottom right panel is the aforementioned count rate rate light curve binned at 60\,s. The bottom left panel is the background subtracted event map, representing the spectra at each time bin of the light curve and every energy. The legend of the event map is in the top right, showing the amount of counts for each color in the bins.

We can see from the dynamic spectrum that the assumption of the counts loss over all wavelengths is not applicable. The Trough looks like a blue triangle in the count map. In ingress and egress only very soft photons are absorbed and deeper into the Trough more and more energetic X-rays are also affected. In the center, photons up to 2\,keV are missing; counts with energies larger than 2\,keV are present even in the center.

\begin{figure*}
    \centering
    \includegraphics[width=\linewidth, trim=40 0 0 0]{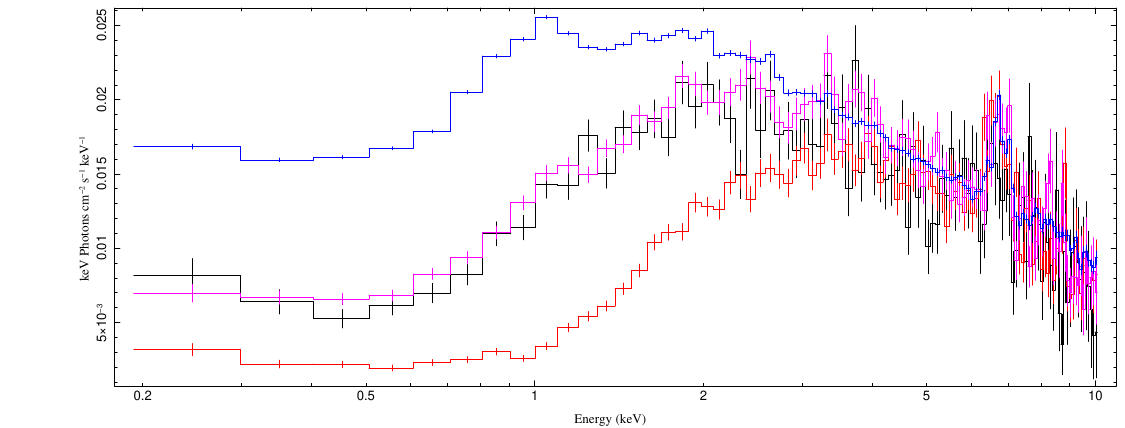}
    \caption{Comparison of ingress (black histogram), trough (red histogram), egress (purple histogram), and outside the trough (blue histogram).}
    \label{fig:TroughComparison}
\end{figure*}

To understand what is happening in the Trough, we can extract spectra at each of the time bins presented in the dynamic spectrum and dynamically combine them. The result of this is shown in Figure~\ref{fig:TroughComparison} where we present four different spectra. The black and purple histograms are the ingress and egress of this event, the red histogram is the bottom of the Trough, and the blue histogram is every point outside the Trough.

It is immediately apparent that this drop in counts corresponded to a smooth, symmetric process since the ingress and egress curves are identical (within uncertainties). We can note how significant the event is as the flux below 1\,keV is an order of magnitude lower than outside this trough. We fit the red histogram\footnote{We did not include the \textit{NuSTAR} data because its exposure started at the end of the egress.} with our cooling flow and DEM model. Since the high energy X-rays (particularly the Fe lines) do not show significant changes, we froze the cooling flow's accretion rate $\dot{M}$, maximum temperature $T_\mathrm{max,1}$, and slope $\beta$. The abundances of the plasma models were also frozen since the PN camera does not have the resolution for constraining line fluxes. Only the column density and covering fraction were free for the cooling flow component.

For the power law DEM, we initially had allowed the maximum temperature $T_\mathrm{max,2}$ and slope $\gamma$ to account for the possibility that the shock between the \Change{accretion halo} and \Change{accretion disk} changed during the Trough. However, these two parameters were completely unconstrained when left free. We then elected to freeze $T_\mathrm{max,2}$ and $\gamma$, so the only free parameters were the column densities $n_\mathrm{H,cflow}$ and $n_\mathrm{H,DEM}$ and the covering fraction $f$. We can interpret this as the physical parameters of X-ray sources have not changed, only the amount of absorption that the X-rays experience. We should note that we also left the power law DEM normalization free, but as will be seen below, the value it takes is of no consequence due to exceptionally high absorption.

With only three absorption parameters, the free parameters were well constrained and quality fits were found. The result of this trough fitting are given in Table~\ref{tab:TroughFit} where we can see that the change in absorption is significant. The power law DEM's column density is especially apparent as it increased by two orders of magnitude, effectively cutting off all of the emission given how weak it already was in comparison to the cooling flow. This is not to say that the cooling flow did not also experience increased absorption as its density increased by 10 times. These increases are consistent with the previous interpretation of dips in the light curves being predominantly due to absorption.

\begin{deluxetable}{lcc}
    \tablecaption{Absorption parameters during and outside the Trough \label{tab:TroughFit}}
    \tablehead{
        \colhead{Parameter} & \colhead{Trough} & \colhead{Outside}
    }
    \startdata
    $n_\mathrm{H, cflow}$ $(10^{22}\,\mathrm{cm}^{-2})$ & $2.55_{-0.04}^{+0.04}$ & $0.22_{-0.01}^{+0.01}$\\
    $f$ & $0.971_{-0.002}^{+0.002}$ & $0.92_{-0.03}^{+0.02}$ \\
    $n_\mathrm{H, DEM}$ $(10^{22}\,\mathrm{cm}^{-2})$ & $1.02_{-0.04}^{+0.04}$ & $0.06_{-0.02}^{+0.04}$\\
    \enddata
\end{deluxetable}

The change in the covering fraction to 97 percent implies that the absorption is much more significant than simply a higher column density. More specifically, it means a further 5 percent of the boundary layer is blocked from our view and explains the dramatic flux change in Figure~\ref{fig:TroughComparison}. The clump occultations from \citet{Hamaguchi2016} can explain such an increase in covering fraction, but it would be difficult for a single clump to block our line of sight for nearly half an hour. Additionally, there are more details in the light curve after the Trough that point to the cause being within the Be disk.

\begin{figure*}
    \centering
    \includegraphics[width=\linewidth]{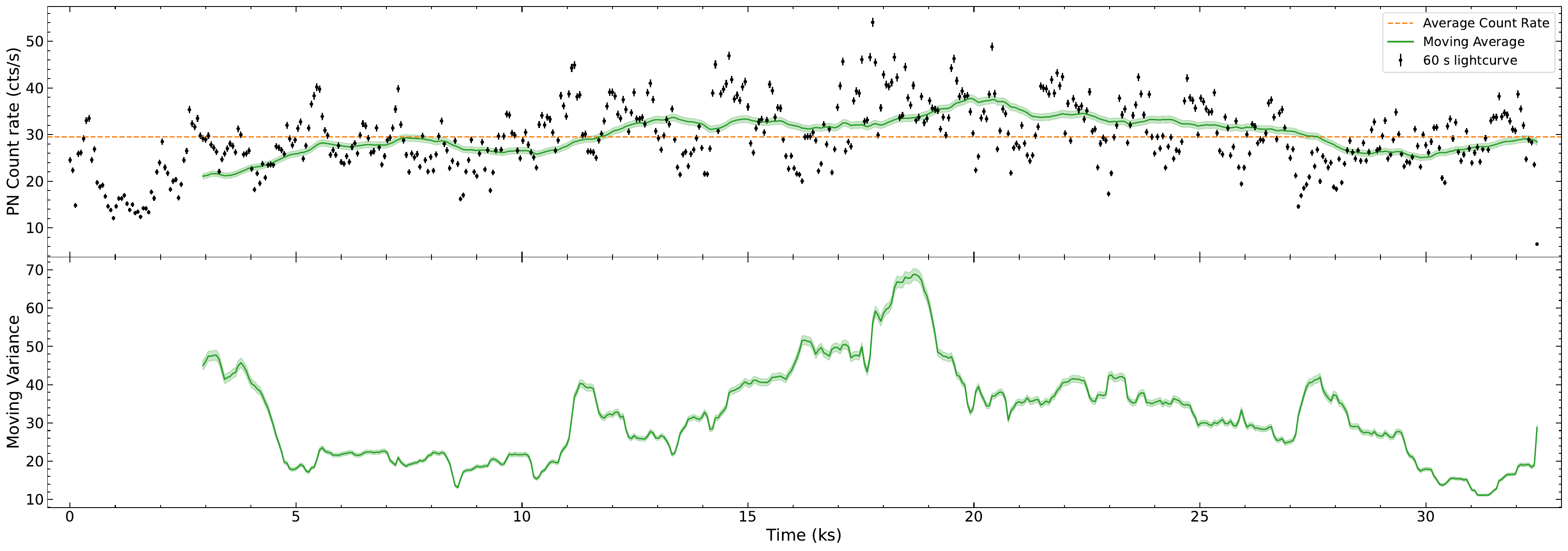}
    \caption{Top: PN light curve of $\gamma$ Cas in 60\,s bins (black dots). Horizontal dashed red line is the average count rate of the light curve. The green line is the moving average using 5 bins (300\,s) at a time. Bottom: Moving variance in \Change{5 bins} (300\,s) as a proxy for variability amplitude.}
    \label{fig:PNMovingAverage}
\end{figure*}

In the top panel of Figure~\ref{fig:PNMovingAverage} we plot the PN light curve in 60\,s bins with the average count rate (red dashed line). We also plot the moving average count rate in green to illustrate broadband changes in the emission from the system. The moving average was over 300\,s (5 bins), so it starts much lower than average because of the Trough event at the beginning of the observation. Approximately two hours after the egress, the moving average starts rising to above average count rates. This increased average count rate lasts for about 4 hours until it returns to a quiescent state. A similar effect appears in the amplitude of the flickering, which we estimate using the moving variance (calculated for the same 300\,s bins) plotted in the bottom panel of Figure~\ref{fig:PNMovingAverage}. Again note that the beginning of the moving variance is biased by the large absorption event, so it should be considered to actually start around 5\,ks. We can see in the moving variance that around the same time as the moving average starts increasing (12\,ks), there is an overall increase in the flickering amplitude. Similarly, the variance decreases with the count rate average.

We have noted throughout this paper that the flickering due to the \Change{accretion disk} instabilities is characteristically over short time scales of order a minute. The overall increase in the count rate occurs over a much longer time scale, so it represents a physical change to the system as more X-rays escape the system. This could correspond with a change in the amount of absorption in the system, but this would not explain the increase in the flickering amplitude. Instead, the system could be producing more X-rays. Given our hypothesis of the two proposed sources, an increase in X-ray production would mean that more mass is going through the shock between the \Change{accretion halo} and \Change{accretion disk} and/or accreting onto the WD.

Be decretion disks are known to have complex structures, such as spiral arm density waves or radially flowing clumps \citep{Okazaki2002,Escolano2015,Cyr2020,Suffak2022}. We propose that the Trough in the \textit{XMM} observation is due to the WD encountering one of these structures. The enhanced density of the structure caused the increase in the column densities of both X-ray sources and increased in the covering fraction. Thus the increase in $f$ is not due to geometric changes, which are unlikely given our viewing angle of the system is 45 degrees \citep{Nemravova2012} meaning we view the boundary layer at the same angle throughout the orbit, but instead a change in the amount of absorbers around the WD.

Some high density gas would have been trapped in the WD's gravitational well and become part of the \Change{accretion halo} and eventually joined the \Change{accretion disk}. This process would take time, explaining why there is not a significant increase in the X-rays right after the Trough in Figure~\ref{fig:PNMovingAverage}. When the higher density gas did reach the \Change{accretion halo-accretion disk} shock, though, it applied increased pressure on the \Change{accretion disk's} outer edge, generating waves in the \Change{accretion disk}. The increase in pressure then propagated down the \Change{accretion disk} till it reached the inner edge and raised the mass accretion rate, thereby the increasing X-rays. One theory for the source of flickering is wave propagation, so changing the amplitude of an existing wave or initializing more waves would correspond to an increase in flickering \citep{Bruch1992}.

Based on the above analysis, we propose to give the Trough the more appropriate name of a deep absorption event to better differentiate it from the more common absorption features that cause \citet{Smith2019}'s \Change{softness dips}. This also allows us to classify these two events more explicitly based on the features noted above. A \Change{deep absorption event} is characterized by a large drop in the observed counts over an extended period with a U-shaped trough in the light curve. The \Change{softness dips} are instead short lived, smaller scale loss of counts that are over much shorter times, giving a V-shape dip in the light curve. The difference in shape corresponds to the physical differences in the causes. A \Change{deep absorption event} is caused by the WD system being enveloped by some high-density structure in the Be disk while the \Change{softness dips} are due to occultations by clumps in the Be wind as proposed by \citet{Hamaguchi2016}.

The properties of the \Change{deep absorption event} also gives arguments against the wind shock interpretation, either between the WD and Be winds from \citet{Hamaguchi2016} or embedded wind shocks of just the Be wind, of the secondary source. In the former case of interactions between two winds, we would see an increase in emission because their would be more gas for the WD wind to interact with. On embedded wind shock possibility, it is difficult to make a physical argument for how the entire wind volume would cease to be shocking or be suddenly more absorbed.

We can estimate the size of the high-density structure if we assume that it has a symmetric, constant distribution of gas. The ``transit time'' of this event will be directly proportional to the size of the WD system and the structure
\begin{equation}
    \frac{R_\mathrm{struct}}{R_\mathrm{AD}} = \frac{t_\mathrm{ingress}}{t_\mathrm{trough}}.
\end{equation}
Earlier we estimated the ingress time to be $t_\mathrm{ingress}\approx 420$\,s and the bottom of the Trough to be $t_\mathrm{trough}\approx840$\,s. This means the high-density structure had a radius of $R_\mathrm{struct}\approx2R_\mathrm{AD}$. These numbers are suspiciously perfect, but as noted before, the flickering makes it difficult to determine the transition points. Regardless of the precision of the chosen transition times, the result will be of order twice the \Change{accretion disk} outer radius.

The size of the high-density structure is notable. The outer radius of a CV's \Change{accretion disk} is approximately $R\approx 100 R_\mathrm{WD}$ \citep{Pringle1985}, so the high-density structure is almost $2 R_\odot$ in radius. From Equation~\eqref{eq:ScaleHeight}, the scale height of the decretion disk at the WD's radius is $H\approx1.3 R_\odot$, bringing into question our spherical assumption. It may be the case that the high-density structure is asymmetric so as to be wider in the $\phi$-direction in the disk.

On the other hand, it could imply that the outer radius of the \Change{accretion disk} is smaller in $\gamma$\,Cas's WD. In a classical CV, the disk expands into a vacuum till it finds a hydrostatic radius \citep{Klement2017}. In contrast, the WD in $\gamma$\,Cas is always surrounded by the Be disk and the \Change{accretion halo}.. The surrounding gas will provide additional inwards pressure, bringing the hydrostatic surface closer to the WD. More sophisticated calculations could use the triangular shape of the \Change{deep absorption event} in Figure~\ref{fig:DynamicSpectra} to calculate densities along different sight lines to reconstruct the size of the structure based on the apparent non-constant density.

\section{Conclusions}\label{sec:Conclusions}

We showed that the X-ray variability in $\gamma$\,Cas shows different signatures between the hard and soft bands. The latter appears to be purely due to absorption, whereas the former shows the same autocorrelation as flickering in non-magnetic CVs. We also analyzed the spectra from simultaneous \textit{XMM} and \textit{NuSTAR} observations within the framework of two different emission sources based on both the timing analysis and similar, albeit less directly interpreted, modeling in the literature. Our model consisted of a cooling flow model to describe the hypothesized WD that is generating the hard X-rays along with a power-law DEM for the unknown, softer secondary source. The best-fit parameters on the cooling flow model were consistent with estimates of the WD properties from optical studies, adding further evidence for the WD hypothesis.

The power law DEM provided reasonably constrained parameters and was shown to reproduce the He-like lines in the RGS that the cooling flow could not. This second source is proposed to be a shock between the in-falling Be decretion disk material trapped by the WD's gravitational well and the WD's accretion disk. The shock is specifically due to the difference in angular momentum between the in falling gas and the disk, producing characteristically softer X-rays than the accretion shock at the base of the accretion disk.

Our estimate of the maximum temperature of this shock $T_\mathrm{max,2}$ was the only under-constrained parameter due to the large positive error. Line fluxes are the primary parameter in the power law DEM model we used (see \citealp{Huenemoerder2020}), so we are limited in the resolution and sensitivity of the RGS at the shorter wavelengths. The inclusion of lines below 6\,\AA\ would significantly improve the power law DEM's confidence intervals.

The upcoming \textit{XRISM} observation\footnote{\url{https://heasarc.gsfc.nasa.gov/docs/xrism/proposals/AO1_ver_final.html}} will provide some of these lines, but we caution the use of this data with non-simultaneous data sets from other telescopes. The variability of $\gamma$\,Cas extends over observation epochs, so non-simultaneous data cannot be fit together. This is evident in our best-fit parameters, which do not match those by \citet{Tsujimoto2018} despite using the same \textit{NuSTAR} data. If \textit{XRISM}'s Gate Valve is opened in the near future, then $\gamma$\,Cas should be proposed again to allow for resolved lines across the 0.2 -- 12\,keV waveband that shows both emission sources. If the Gate Valve remains closed, then joint proposals between \textit{XMM} and \textit{XRISM} will be necessary.

Finally, we analyzed a unique feature of the \textit{XMM} dataset, which we call a deep absorption event, that shows a remarkable loss of half the expected count rate over an extended period of almost 2\,ks. The exact cause of the \Change{deep absorption event} is not known, but we argue that it is evidence of the WD encountering some form of decretion disk instability/structure that enveloped the WD and its \Change{accretion disk} and significantly increased the absorption. Afterwards, there is a commensurate rise in the count rate, possibly due to extra material entering the accretion system and pushing on the \Change{accretion disk}.

A question of interest is the frequency of \Change{deep absorption events}. If the \Change{deep absorption event} was caused by a spiral arm in the Be disk, there will be a periodicity like the interactions of the neutron star and gas stream in GX\,301-2 \citep{Leahy2008}. One-arm spiral arm waves have previously been detected in $\gamma$\,Cas \citep{Berio1999}, but subsequent measurements found that this arm disappeared \citep{Stee2012}.

On the other hand, if the WD encountered a clump in the disk, it will be a stochastic event with no periodicity. With only one example of a \Change{deep absorption event}, we cannot make any statements of periods. There are potentially more of these events in Figure~\ref{fig:PNMovingAverage}, such as between 15 -- 20 ks, but the enhanced flickering makes it difficult to pinpoint the U-shape characteristic of the \Change{deep absorption event}. \textit{NICER} has observed $\gamma$\,Cas across approximately 2 consecutive orbital periods. As a timing-focused instrument sensitive to softer X-rays, \textit{NICER} is well equipped to answer the question of \Change{deep absorption event} frequencies. If a \Change{deep absorption event} is found at the same orbital phase, then it may be evidence of a new one-arm wave in the Be disk.

This work shows that the WD hypothesis for $\gamma$\,Cas's X-ray emission can explain both spectral and temporal features. Yet simple accretion X-ray generation is not enough to fully explain what we see, adding further complications and mysteries to untangle. $\gamma$\,Cas and its analogues should continue to be monitored with a wide range of instruments and wavelengths to find more cases of the features we have discussed here.


\begin{acknowledgements}

Support for SJG, DPH, HMG, and NS was provided by NASA through the Smithsonian
Astrophysical Observatory (SAO) contract SV3-73016 to MIT for Support
of the Chandra X-Ray Center (CXC) and Science Instruments. CXC is
operated by SAO for and on behalf of NASA under contract NAS8-03060. SJG and HMG also received support by NASA through Chandra Award Number GO3-24005X issued by the CXC.

This research has made use of ISIS functions (ISISscripts) provided by
ECAP/Remeis observatory and MIT (\url{http://www.sternwarte.uni-erlangen.de/isis/}).

We thank M. Suffak and T. H. de Amorim for their insights into Be disk structure for putting our data into context.

\end{acknowledgements}

\facility{ \textit{CXO} (HETG/ACIS), \textit{XMM-Newton} (PN, RGS), \textit{NuSTAR}}

\software{\textsc{CIAO} \citep{Fruscione2006}, \textsc{ISIS} \citep{Houck2000}, \textsc{SAS} \citep{Gabriel2004}}

\appendix 

\section{Tables of Observation Identification Numbers}\label{sec:ObsIDs}

\begin{deluxetable}{lllc}
    \tablecaption{$\gamma$ Cas Observation IDs.\label{tab:ObsIds}}
    \tablehead{
        \colhead{Observatory} & \colhead{Obs Id} & \colhead{Start Date} & \colhead{Exposure time (ks)}
    }
    \startdata
        \textit{Chandra} & 1895 & 2001-08-10 & 51.47 \\
        \textit{XMM} & 0743600101\tablenotemark{a} & 2014-07-24 & 34 \\
        \textit{NuSTAR} & 30001147002\tablenotemark{a} & 2014-07-24 & 30.76 \\
    \enddata
    \tablenotetext{a}{Simultaneous Observation}
\end{deluxetable}

\begin{deluxetable}{llcc}
  \tablecaption{Observational Information for other star's HETGS Data \label{tab:OtherObsIds}}
  \tablehead{
    \colhead{Star} &
    \colhead{Obs Id} &
    \colhead{Start Date} &
    \colhead{Exposure time (ks)}
  }
  \startdata
  $\zeta$ Pup & 640& 2000-03-28& 67.73\\
   & 21113& 2018-07-01& 17.72\\
   & 21112& 2018-07-02& 29.70\\
   & 20156& 2018-07-03& 15.51\\
   & 21114& 2018-07-05& 19.69\\
   & 21111& 2018-07-06& 26.86\\
   & 21115& 2018-07-07& 18.09\\
   & 21116& 2018-07-08& 43.39\\
   & 20158& 2018-07-30& 18.41\\
   & 21661& 2018-08-03& 96.88\\
   & 20157& 2018-08-08& 76.43\\
   & 21659& 2018-08-22& 86.34\\
   & 21673& 2018-08-24& 14.95\\
   & 20154& 2019-01-25& 46.97\\
   & 22049& 2019-02-01& 27.69\\
   & 20155& 2019-07-15& 19.69\\
   & 22278& 2019-07-16& 30.51\\
   & 22279& 2019-07-17& 26.05\\
   & 22280& 2019-07-20& 25.53\\
   & 22281& 2019-07-21& 41.74\\
   & 22076& 2019-08-01& 75.12\\
   & 21898& 2019-08-17& 55.70\\
  Capella & 1103 & 1999-09-24 & 40.52 \\
  EX Hya & 7449 & 2007-05-13 & 128.54 \\
  SS Cyg & 646 & 2000-08-24 & 47.34
  \enddata
\end{deluxetable}

\begin{deluxetable}{rrc|rrc}
  \tablecaption{Observational Information for $\theta^1\,$Ori~C HETGS Data \label{tab:t1ocobs}}
  \tablehead{
    \colhead{Obs Id} &
    \colhead{Start Date} &
    \colhead{Exposure time (ks)} &
    \colhead{Obs Id} &
    \colhead{Start Date} &
    \colhead{Exposure time (ks)}
  }\label{tbl:oriobs}
  \startdata
      3& 1999-10-31& 49.52&    23115& 2020-01-04& 29.67\\
      4& 1999-11-24& 30.91&    22335& 2020-01-06& 29.67\\
   2567& 2001-12-28& 46.36&    23005& 2020-01-08& 24.73\\
   2568& 2002-02-19& 46.33&    23120& 2020-01-11& 39.54\\
   7407& 2006-12-03& 24.63&    23012& 2020-04-01& 10.81\\
   7410& 2006-12-06& 13.07&    23206& 2020-04-02& 17.71\\
   7408& 2006-12-19& 24.86&    23207& 2020-04-04& 14.75\\
   7409& 2006-12-23& 27.09&    23208& 2020-04-05& 14.75\\
   8568& 2007-08-06& 35.86&    23011& 2020-04-21& 51.69\\
   8589& 2007-08-08& 50.40&    22341& 2020-04-29& 32.12\\
   8897& 2007-11-15& 23.64&    23233& 2020-05-01& 34.59\\
   8896& 2007-11-30& 22.66&    23010& 2020-07-27& 25.72\\
   8895& 2007-12-07& 24.85&    23001& 2020-07-28& 25.62\\
  23008& 2019-11-27& 47.43&    23009& 2020-07-28& 25.01\\
  22893& 2019-12-02& 24.73&    22340& 2020-10-14& 25.62\\
  22994& 2019-12-05& 24.73&    24832& 2020-10-15& 27.59\\
  23087& 2019-12-08& 39.54&    22997& 2020-10-15& 26.60\\
  22904& 2019-12-10& 36.58&    24834& 2020-10-16& 26.91\\
  23097& 2019-12-11& 35.88&    22342& 2020-10-20& 34.50\\
  22337& 2019-12-13& 37.66&    24842& 2020-10-21& 29.57\\
  23006& 2019-12-14& 24.73&    22993& 2020-10-23& 24.63\\
  22343& 2019-12-15& 24.73&    22998& 2020-11-01& 23.09\\
  23003& 2019-12-21& 24.74&    22999& 2020-11-08& 35.58\\
  23104& 2019-12-21& 24.73&    24830& 2020-11-22& 26.52\\
  22336& 2019-12-22& 25.59&    24622& 2020-11-23& 24.56\\
  23007& 2019-12-24& 37.41&    24873& 2020-11-24& 24.74\\
  22339& 2019-12-26& 31.64&    24874& 2020-11-24& 25.72\\
  22892& 2019-12-26& 30.65&    24829& 2020-11-27& 26.45\\
  22995& 2019-12-27& 38.74&    24623& 2020-11-29& 24.74\\
  22338& 2019-12-30& 39.15&    24624& 2020-12-09& 29.67\\
  22334& 2019-12-31& 24.73&    23002& 2020-12-10& 30.66\\
  23000& 2020-01-01& 42.50&    23004& 2020-12-12& 32.14\\
  22996& 2020-01-03& 26.70&    24831& 2020-12-25& 30.66\\
  23114& 2020-01-03& 37.56&    24906& 2020-12-25& 28.68\\
  \enddata
\end{deluxetable}


\clearpage

\bibliography{bib}{}
\bibliographystyle{aasjournal}

\end{document}